\documentclass{sig-alternate-05-2015}

\usepackage[
  pass,
]{geometry}

\newfont{\mycrnotice}{ptmr8t at 7pt}
\newfont{\myconfname}{ptmri8t at 7pt}
\let\crnotice\mycrnotice%
\let\confname\myconfname%

\usepackage[ruled,vlined]{algorithm2e}
\usepackage{booktabs}
\usepackage{calc}
\usepackage[inline,shortlabels]{enumitem}
\usepackage{graphicx}
\usepackage{multirow}
\usepackage[caption=false]{subfig}
\usepackage{times}
\usepackage{url}
\usepackage{xspace}
\usepackage{verbatim}
\usepackage{amsmath}
\usepackage{amssymb}
\usepackage{pifont}
\usepackage{float}

\makeatletter
\def\BState{\State\hskip-\ALG@thistlm}
\makeatother

\newcommand{\urlwofont}[1]{ \urlstyle{same}\url{#1} }
\newcommand{\para}[1]{\noindent\textbf{#1}.}
\renewcommand{\vec}[1]{\mathbf{#1}}

\begin{document}

\permission{Permission to make digital or hard copies of part or all of this work for
personal or classroom use is granted without fee provided that copies are
not made or distributed for profit or commercial advantage and that copies
bear this notice and the full citation on the first page. Copyrights for
third-party components of this work must be honored. For all other uses,
contact the owner/author(s).}
\conferenceinfo{Neu-IR '16 SIGIR Workshop on Neural Information Retrieval}{July 21, 2016, Pisa, Italy \\
{\mycrnotice{\copyright 2016 Copyright held by the owner/author(s).}}}

\title{Using Word Embeddings for Automatic Query Expansion}


\numberofauthors{4}
\author{
\alignauthor
Dwaipayan Roy \\
       \affaddr{CVPR Unit}\\
       \affaddr{Indian Statistical Institute} \\
       \affaddr{Kolkata, India}\\
       \email{dwaipayan\_r@isical.ac.in}
\and
\alignauthor
Debjyoti Paul\\
       \affaddr{CVPR Unit}\\
       \affaddr{Indian Statistical Institute} \\
       \affaddr{Kolkata, India}\\
       \email{debjyoti93.paul@gmail.com}
\and 
\alignauthor
Mandar Mitra\\
       \affaddr{CVPR Unit}\\
       \affaddr{Indian Statistical Institute} \\
       \affaddr{Kolkata, India}\\
       \email{mandar@isical.ac.in}
\and 
\alignauthor
Utpal Garain\\
       \affaddr{CVPR Unit}\\
       \affaddr{Indian Statistical Institute} \\
       \affaddr{Kolkata, India}\\
       \email{utpal@isical.ac.in}}

\maketitle

\begin{abstract}
In this paper a framework for Automatic Query Expansion (AQE) is proposed using distributed neural language model word2vec. Using semantic and contextual relation in a distributed and unsupervised framework, word2vec learns a low dimensional embedding for each vocabulary entry. 
Using such a framework, we devise a query expansion technique, where related terms to a query are obtained by K-nearest neighbor approach. 
We explore the performance of the AQE methods, with and without feedback query expansion, and a variant of simple K-nearest neighbor in the proposed framework.
Experiments on standard TREC ad-hoc data (Disk 4, 5 with query sets 301-450, 601-700) and web data (WT10G data with query set 451-550) shows 
significant improvement over standard term-overlapping based retrieval methods.
However the proposed method fails to achieve comparable performance with statistical oc-occurrence based feedback method such as RM3.
We have also found that the word2vec based query expansion methods perform similarly with and without any feedback information.
\end{abstract}

\section{Introduction}
\label{sec:intro}
In recent times, the IR and Neural Network (NN) communities have started to
explore the application of deep neural network based techniques to various
IR problems. A few studies have focused in particular on the use of
\emph{word embeddings} generated using deep NNs.
A word embedding is a mapping that associates each word or phrase occurring
in a document collection to a vector in $\mathbb{R}^n$, where $n$ is
significantly lower than the size of the vocabulary of the document
collection. If $a$ and $b$ are two words, and $\vec{a}$ and $\vec{b}$ are
their embeddings, then it is expected that the distance between $\vec{a}$
and $\vec{b}$ is a quantitative indication of the semantic relatedness
between $a$ and $b$. Various different techniques for creating word
embeddings --- including Latent Semantic Analysis
(LSA)~\cite{deerwester1990indexing} and probabilistic
LSA~\cite{hofmann1999probabilistic}
--- have been in use for many years. However, interest in the use of word
embeddings has been recently rekindled thanks to work by Mikolov et
al.~\cite{Mikolov13}, and the availability of the
\emph{word2vec}\footnote{https://code.google.com/p/word2vec/} software
package.
It has been reported that the semantic relatedness between words is
generally accurately captured by the vector similarity between the
corresponding embeddings produced by this method. Thus, this method
provides a convenient way of finding words that are semantically related to
any given word.

Since the objective of Query Expansion (QE) is to find words that are
semantically related to a given user query, it should be possible to
leverage word embeddings in order to improve QE effectiveness. Let $Q$ be a
given user query consisting of the words $q_1, q_2, \ldots, q_m$. Let
$w^{(i)}_1, w^{(i)}_2, \ldots, w^{(i)}_k$ be the $k$ nearest neighbours
(kNN) of $q_i$ in the embedding space. Then, these $w^{(i)}_j$s constitute
a set of obvious candidates from which terms may be selected and used to
expand $Q$. Of course, instead of considering terms that are proximate
neighbours of individual query words, it is generally preferable
to consider terms that are close to the query as a whole.\footnote{This
  idea has been used in a number of traditional, effective QE techniques,
  e.g., LCA~\cite{xu_query} and RM3~\cite{rm3}. In these techniques, 
  expansion terms are selected on the basis of their association with all query terms.} 


While word embeddings have been shown to be useful in some specialised
applications (e.g., clinical decision support~\cite{GoodwinH14} and
sponsored search~\cite{GrbovicDRSB15}) and for cross-lingual
retrieval~\cite{VulicM15}, the obvious way of using embeddings for QE seems
not to have been explored within the standard ad hoc retrieval task
setting. Our goal in this work is to study how word embeddings may be
applied to QE for ad hoc retrieval. Specifically, we are looking for
answers to the following questions.
\begin{enumerate}
\item Does QE, using the nearest neighbours of query terms, improve retrieval
  effectiveness?
\item If yes, is it possible to characterise the queries for which this QE
  method does / does not work?
\item How does embedding based QE perform compared to an established QE
  technique like RM3~\cite{rm3}?
\end{enumerate}
We try a few different embedding based QE methods. These methods are
described in more detail in the next section. Experiments on a number of
TREC collections (Section~\ref{sec:eval}) show that these QE methods
generally yield significant improvements in retrieval effectiveness when
compared to using the original, unexpanded queries. However, they are all
significantly inferior to RM3. We discuss these results in greater detail
in Section~\ref{sec:discuss}. Section \ref{sec:fw} concludes the paper.


%
%

\section{Word Embedding based Query Expansion}
\label{sec:method}
In this section, we first describe three QE methods using the individual embeddings of the terms.
The first method is a simple, kNN based QE method that
makes use of the basic idea outlined in Section \ref{sec:intro}. 
Unlike pseudo relevance feedback (PRF) based QE methods, this method does not require an
initial round of retrieval. 
The second approach we tried is a straightforward variation of the first approach that uses word embeddings
in conjunction with a set of pseudo relevant documents. 
In the third method, we propose an approach that is inspired by \cite{weston2014memory}. 
In this approach, the nearest neighbours are computed in an incremental fashion as elaborated below. 
Next, we describe how we obtain an extended query term set by using compositionality of terms.
In all our methods, we used \textit{word2vec}\cite{Mikolov13} for computing word embeddings.

%

\subsection{Pre-retrieval kNN based approach} 
\label{knn-pre}
Let the given query $Q$ be $\{q_1, \ldots, q_m\}$. In this simple approach, 
we define the set $C$ of candidate expansion terms as 
\begin{equation}
  C = \bigcup_{\vec{q}\in Q} \mathit{NN}(\vec{q})   \label{eq:qe1}
\end{equation}
where $\mathit{NN}(\vec{q})$ is the set of $K$ terms that are closest to $q$ in
the embedding space.\footnote{Bold-faced notation $\vec{w}$ denotes the embedded vector corresponding to a word $w$}
For each candidate expansion term $t$ in $C$, we compute the mean cosine
similarity between $t$ and all the terms in $Q$ following Equation \ref{eq:sim}. 
\begin{equation}
    \mathit{Sim}(t, Q) = \frac{1}{|Q|}\sum_{q_i\in Q}{\vec{t}.\vec{q_i}}    \label{eq:sim}
\end{equation}

The terms in $C$ are sorted on the basis of this mean score, 
and the top $K$ candidates are selected as the actual expansion terms.

\subsection{Post-retrieval kNN based approach} 
\label{knn-post}
In our next approach, we use a set of pseudo-relevant documents (PRD) ---
documents that are retrieved at top ranks in response to the initial query
--- to restrict the search domain for the candidate expansion terms.
Instead of searching for nearest neighbours within the entire vocabulary of
the document collection, we consider only those terms that occur within
PRD. The size of PRD may be varied as a parameter. The rest of the
procedure for obtaining the expanded query is the same as in
Section~\ref{knn-pre}.

\subsection{Pre-retrieval incremental kNN based approach} 
\label{knn-pre-incre}
The incremental nearest neighbour method is a simple extension of the
pre-retrieval kNN method that is based on~\cite{weston2014memory}. Instead
of computing the nearest neighbours for each query term in a single step,
we follow an incremental procedure. The first assumption in this method is
that, the most similar neighbours have comparatively lower drift than the
terms occurring later in the list in terms of similarity. Since the most
similar terms are the strongest contenders for becoming the expansion
terms, it may be assumed that these terms are also similar to each other,
in addition to being similar to the query term. Based on the above
assumption, we use an iterative process of pruning terms from
$\mathit{NN}(q)$, the list of candidates obtained for each term $q$ in
EQTS.

We start with $\mathit{NN}(q)$. Let the nearest neighbours of $q$ in order
of decreasing similarity be $t_1, t_2, \ldots, t_N$. We prune the $K$ least
similar neighbours to obtain $t_1, t_2, \ldots, t_{N-k}$. Next, we consider
$t_1$, and reorder the terms $t_2, \ldots, t_{N-k}$ in decreasing order of
similarity with $t_1$. Again, the $K$ least similar neighbours in the
reordered list are pruned to obtain $t'_2, t'_3, \ldots, t'_{N-2k}$. Next,
we pick $t'_2$ and repeat the same process. This continues for $l$
iterations. At each step, the nearest neighbours list is reordered based on
the nearest neighbour obtained in the previous step, and the set is pruned.
Essentially, by following the above procedure, we are constraining the
nearest neighbours to be similar to each other in addition to being similar
to the query term. A high value of $l \geq 10$ may lead to query drift. A
low value of $l \leq 2$ essentially performs similar to the basic
pre-retrieval model. We empirically choose $l = 5$ as the number of
iterations for this method. Let $\mathit{NN}_l(q)$ denote the iteratively
pruned nearest neighbour list for $q$. The expanded query is then
constructed as in Section~\ref{knn-pre}, except that $\mathit{NN}_l(q)$ is
used in place of $\mathit{NN}(q)$ in Equation~\ref{eq:qe1}.

\subsection{Extended Query Term Set}
\label{subsubsec:compose}
Considering NNs of individual query word makes a generalization towards the process of choosing expansion terms
since a single term may not reflect the information need properly.
For example, consider the TREC query \emph{Orphan Drugs} where the respective terms may have multiple associations, not related to the 
actual information need.
The conceptual meaning of conposition of two or more words can be achieved by simple addition of the constituent vectors.

Given a query $Q$ consisting of $m$ terms $\{q_1,\ldots,q_m\}$, we first
construct $Q_c$, the set of query word bigrams.
\[ 
   Q_c = \{ \langle q_1, q_2 \rangle, \langle q_2, q_3 \rangle, \ldots,
             \langle q_{m-1}, q_m \rangle \}
\]
We define the embedding for a bigram $\langle q_i, q_{i+1} \rangle$ as
simply $\vec{q_i} + \vec{q}_{i+1}$, where $\vec{q_i}$ and $\vec{q}_{i+1}$
are the embeddings of words $q_i$ and $q_{i+1}$. Next, we define an
extended query term set (EQTS) $Q'$ as 
\begin{equation}
    Q' = Q \cup Q_c    \label{eq:qe2}
\end{equation}

For the proposed approaches, the effect of compositionality can be integrated by considering $Q'$ of 
Equation \ref{eq:qe2} in place of $Q$ in Equation \ref{eq:qe1} and \ref{eq:sim}.

\subsection{Retrieval}
\label{subsec:ret}
For our retrieval experiments, we used Language Model with Jelinek Mercer
smoothing~\cite{zhai}. The query model for the expanded query is given by 
\begin{equation}
  \label{eq:2}
  P(w|Q_{\mathit{exp}}) = \alpha P(w|Q) ~+~ 
  (1-\alpha) \frac{\mathit{Sim}(w, Q)}{\sum_{w \in Q_{\mathit{exp}}} \mathit{Sim}(w, Q)}
\end{equation}
where $Q_{exp}$ is the set of top $K$ terms from $C$, the set of candidate expansion terms.
As described in Section \ref{subsubsec:compose}, we can use $Q$ or $Q'$ in Equation \ref{eq:2}.
The expansion term weights are assigned by normalizing
the expansion term score (mean similarity with respect to all the terms in
EQTS) by the total score obtained by summing over all top $K$ expansion terms. 
$\alpha$ is the interpolation parameter to use the likelihood estimate of a term in the query, 
in combination with the normalized vector similarity with the query.

\section{Evaluation}
\label{sec:eval}

\begin{table}[!ht]
\small
  \caption{\label{tab:trec}Dataset Overview}
  \centering

\begin{tabular}{@{}l@{~~}c@{~~}c@{~~}l@{~~}c@{~~}c@{~~}c@{~~}}
\toprule
Document & Document & \#Docs & Query Set & Query Ids & Avg qry & Avg \# \\
Collection & Type & & & & length & rel docs \\
\midrule
\multirow{3}{*}{TREC} & \multirow{4}{*}{News} & \multirow{4}{*}{528,155} & TREC 6
& 301-350 & 2.48 & 92.2 \\ 
& & & TREC 7 & 351-400 & 2.42 & 93.4 \\ 
\multirow{1}{*}{Disks 4, 5} & & & TREC 8 & 401-450 & 2.38 & 94.5 \\ 
& & & TREC Robust & 601-700 & 2.88 & 37.2 \\
\midrule
WT10G & Web pages & 1,692,096 & TREC 9-10& 451-550 & 4.04 & 59.7 \\
\bottomrule
\end{tabular}
\end{table}

We explored the effectiveness of our proposed method on the standard ad-hoc 
task using TREC collection as well as on the TREC web collection. 
Preciously, we use the documents from TREC disk 4 and 5 with the query sets TREC 6, 7, 8 and Robust. 
For the web collection, we use WT10G collection.
The overview of the dataset used is presented in Table \ref{tab:trec}. 
We implemented our method~\footnote{Available from \url{https://github.com/dwaipayanroy/QE_With_W2V}} using the 
Apache licensed Lucene search engine\footnote{\url{https://lucene.apache.org/core/}}.
We used the Lucene implementation of the standard language model with linear 
smoothing~\cite{zhai}.

\subsection{Experimental Setup}
\label{subsec:exp-setup}
\para{Indexing and Word Vector Embedding}
\label{subsec:vec-learn}
At the time of indexing of the test collection, we removed the stopwords
following the SMART\footnote{\url{ftp://ftp.cs.cornell.edu/pub/smart/}} stopword-list.
Porter stemmer is used for stemming of words. The stopword removed and stemmed 
index is then dumped as raw text for the purpose of training the neural 
network of \emph{Word2Vec} framework. The vectors are embedded in an abstract 
200 dimensional space with negative sampling using 5 word window on continuous 
bag of words model. 
For the training, we removed any words that appear less than three times in the whole corpus.
These are as par the parameter setting prescribed in~\cite{Mikolov13}.  \\

\para{Parameter setting}
In all our experiments, we only use the `title' field of the TREC topics
as queries.
The linear smoothing parameter $\lambda$ was empirically set to $0.6$, which is producing the optimal results,
after varying it in the range $[0.1, 0.9]$. 
The proposed methods have two unique parameters associated with them; 
$K$, that is the number of expansion terms choosen from $Q_{exp}$ for QE, and
the interpolation parameter $\alpha$.
In addition, the feedback based method (Section \ref{knn-post}) has one more parameter, 
the number of documents to use for feedback.
To compare the best performance of the proposed methods, we explored all parameter grids to find out 
the best performance of the individual approaches.
The corresponding parameters, which are producing the optimal results, are reported in Table \ref{tab:results}
along with the evaluation metrics.

\subsection{Results}
\label{result}

As an early attempt, we compared the effect of applyting composition,
when computing the similarity between an expansion term and the query,
for the pre-retrieval kNN based approach (Section \ref{knn-pre}).
The relative performance is presented in Table \ref{tab:comparison}.
It is clear from the result that applying composition indeed affects the
performance positively.
Hence, we applied composition (for the similarity computation) in the rest of the approaches.

Table \ref{tab:results} shows the performance of the proposed method,
compared with the baseline LM model and feedback model RM3 \cite{rm3}. It
can be seen that the QE methods based on word embeddings almost always
outperform the LM baseline model (often significantly). There does not
seem to be a major difference in performance between the three variants,
but the incremental method seems to be the most consistent in producing
improvements. 
However, the performance of RM3 is significantly superior for all the query sets.

\begin{table}[t]
\small
\centering
\begin{tabular}{@{}l@{~~}c@{~~}l@{~~}l@{~}l@{~~}c@{~~}l@{~~}}
\toprule
& wvec & TREC6 & TREC7 & TREC8 & ROBUST & WT10G \\
& compo & & & & & \\
\midrule
LM      & -   & 0.2303 & 0.1750 & 0.2373 & 0.2651 & 0.1454 \\
Pre-ret & no  & 0.2311 & 0.1800 & 0.2441 & 0.2759 & 0.1582 \\
Pre-ret & yes & 0.2406 & 0.1806 & 0.2535 & 0.2842 & 0.1718 \\

\bottomrule 
\end{tabular}
\caption{Comparison of performance (MAP) between, when only raw query terms are used for finding out NNs 
(using $Q$ in Equation \ref{eq:2}) and when composition is applied (using $Q'$ in Equation \ref{eq:2}).}
\label{tab:comparison}
\end{table}

%
%
%
%
%
%
%

\begin{table}[t]
\small
\centering
\begin{tabular}{@{}l@{~~}l@{~~}c@{~~}c@{~~}c@{~~~}l@{~}l@{~}l@{}}
\toprule
Query & Method & \multicolumn{3}{c}{Parameters} & \multicolumn{3}{c}{Metrics} \\
\cmidrule(r){3-5}
\cmidrule(r){6-8}
& & K & \#fdbck-docs & $\alpha$ & MAP & GMAP & P@5 \\
\midrule

\multirow{5}{*}{TREC 6} 
& LM        & - & - & -         & 0.2303  & 0.0875 & 0.3920 \\
& Pre-ret   & - & 100 & 0.55    & 0.2406* & 0.1026 & 0.4000 \\
& Post-ret  & 30 & 110 & 0.6    & 0.2393 & 0.1028 & 0.4000 \\
& Increm.   & - & 90 & 0.55     & 0.2354 & 0.0991 & 0.4160 \\
& RM3       & 30 & 70 & -       & 0.2634$^{k,p,i}$ & 0.0957 & 0.4360 \\     
\midrule

\multirow{5}{*}{TREC 7}
& LM        & - & - & -         & 0.1750 & 0.0828 & 0.4080 \\
& Pre-ret   & - & 120 & 0.6     & 0.1806 & 0.0956 & 0.4000 \\
& Post-ret  & 30 & 120 & 0.6    & 0.1806* & 0.0956 & 0.4280 \\
& Increm.   & - & 70 & 0.55     & 0.1887* & 0.1026 & 0.4360 \\
& RM3       & 20 & 70 & -       & 0.2151$^{k,p,i}$ & 0.1038 & 0.4160 \\     
\midrule

\multirow{5}{*}{TREC 8} 
& LM        & - & - & -         & 0.2373 & 0.1318 & 0.4320 \\
& Pre-ret   & - & 120 & 0.65    & 0.2535* & 0.1533 & 0.4680 \\
& Post-ret  & 30 & 90 & 0.65    & 0.2531* & 0.1529 & 0.4600 \\
& Increm.   & - & 120 & 0.65    & 0.2567* & 0.1560 & 0.4680 \\
& RM3       & 20 & 70 & -       & 0.2701$^{k,p,i}$ & 0.1543 & 0.4760 \\     
\midrule

\multirow{5}{*}{Robust} 
& LM        & - & - & -      & 0.2651 & 0.1710 & 0.4424 \\
& Pre-ret   & - & 90 & 0.65  & 0.2842* & 0.1869 & 0.4949 \\
& Post-ret  & 30 & 100 & 0.6 & 0.2885* & 0.1901& 0.5010 \\
& Increm.   & - & 90 & 0.6   & 0.2956* & 0.1972 & 0.5051 \\
& RM3       & 20 & 70 & -    & 0.3304$^{k,p,i}$ & 0.2177 & 0.4949 \\     
\midrule

\multirow{5}{*}{WT10G} 
& LM        & - & - & -         & 0.1454 & 0.0566 & 0.2525 \\
& Pre-ret   & - & 80 & 0.6      & 0.1718* & 0.0745 & 0.2929 \\
& Post-ret  & 30 & 90 & 0.6     & 0.1709* & 0.0769 & 0.3071 \\
& Increm.   & - & 100 & 0.55    & 0.1724* & 0.0785 & 0.3253 \\
& RM3       & 20 & 70 & -       & 0.1915$^{k,p,i}$ & 0.0782 & 0.3273 \\     

\bottomrule 
\end{tabular}
\caption{MAP for baseline retrieval and various QE strategies. A * in the
    kNN and Increm.\ columns denotes a significant improvement over the
    baseline. A $k$, $i$, and $p$ in the RM3 column denotes a significant
    improvement over the kNN, Incremental and Post-retrieval QE techniques.
    Significance testing has been performed using paired t-test with 95\% confidence.
\label{tab:results}}
\end{table}

\begin{figure*}
  \centering
  \includegraphics[scale=0.4]{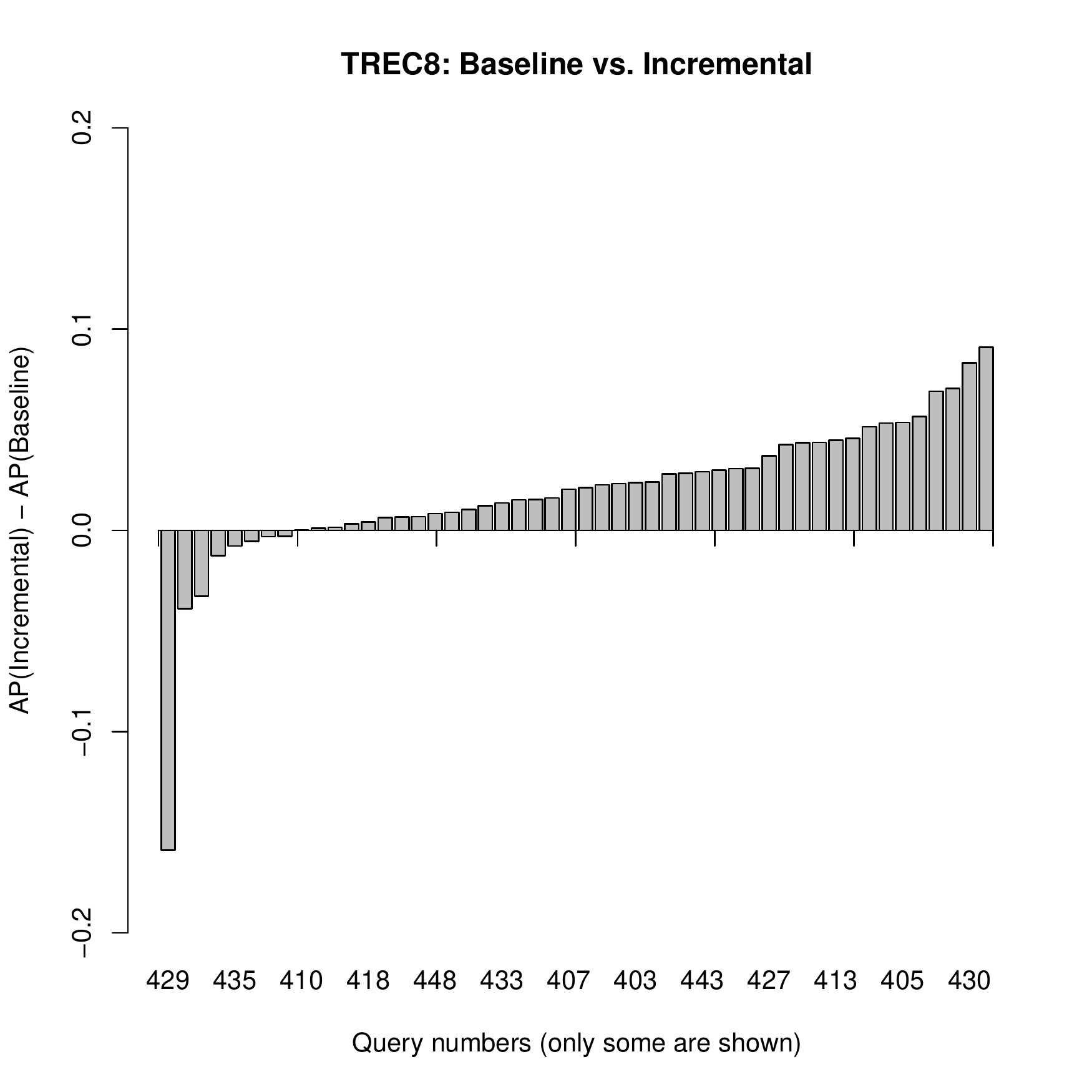}
  \hspace{1cm}
  \includegraphics[scale=0.4]{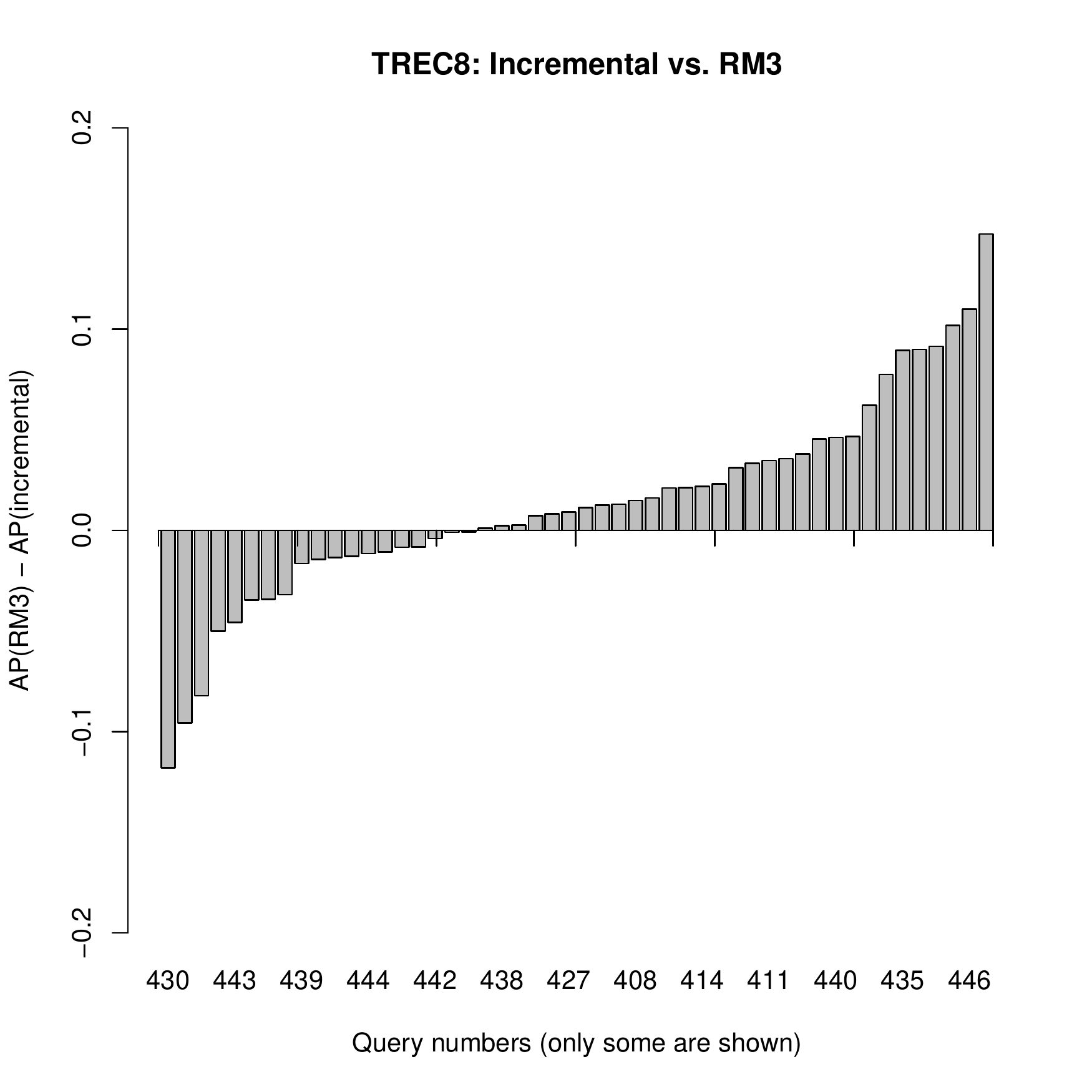}

  \includegraphics[scale=0.4]{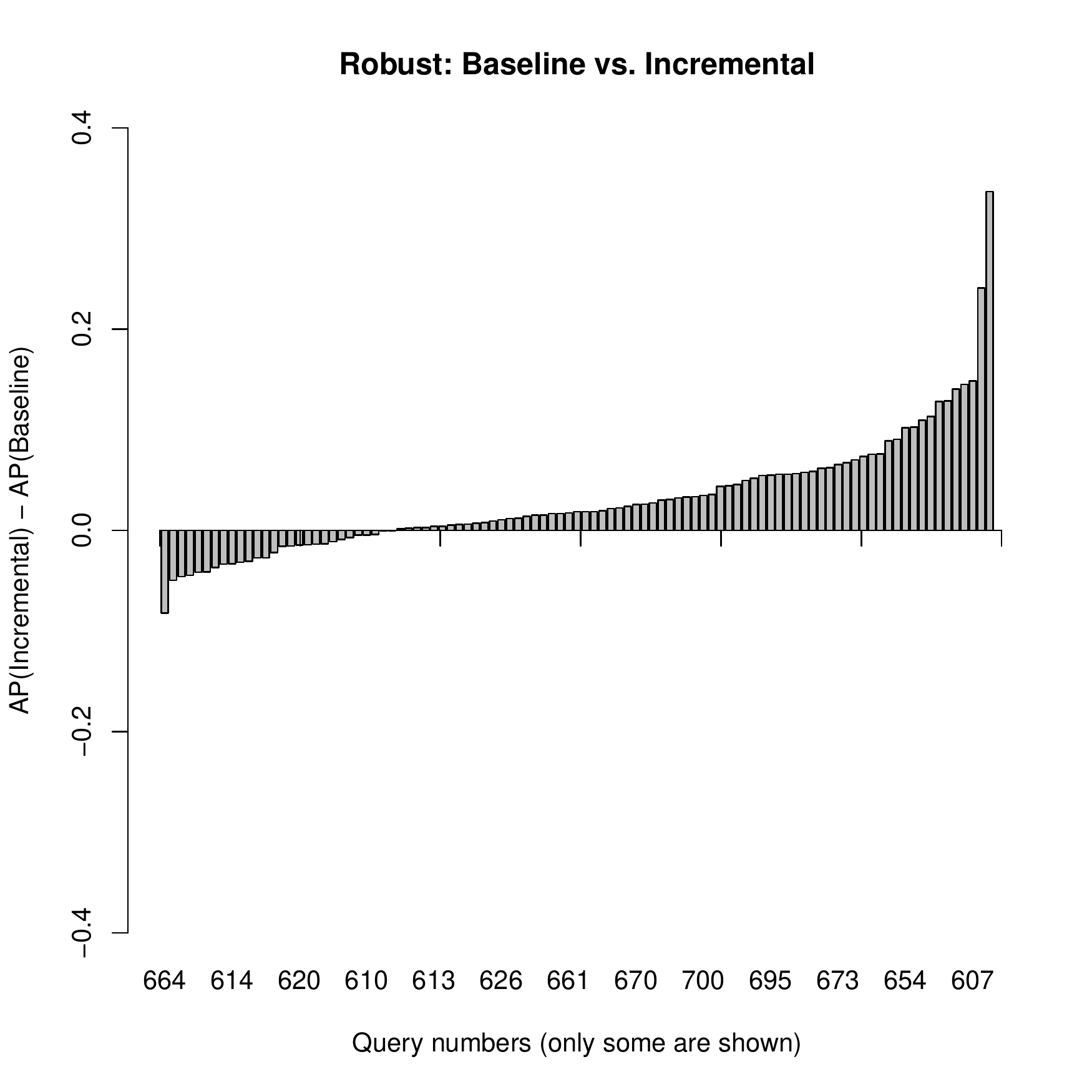}
  \hspace{1cm}
  \includegraphics[scale=0.4]{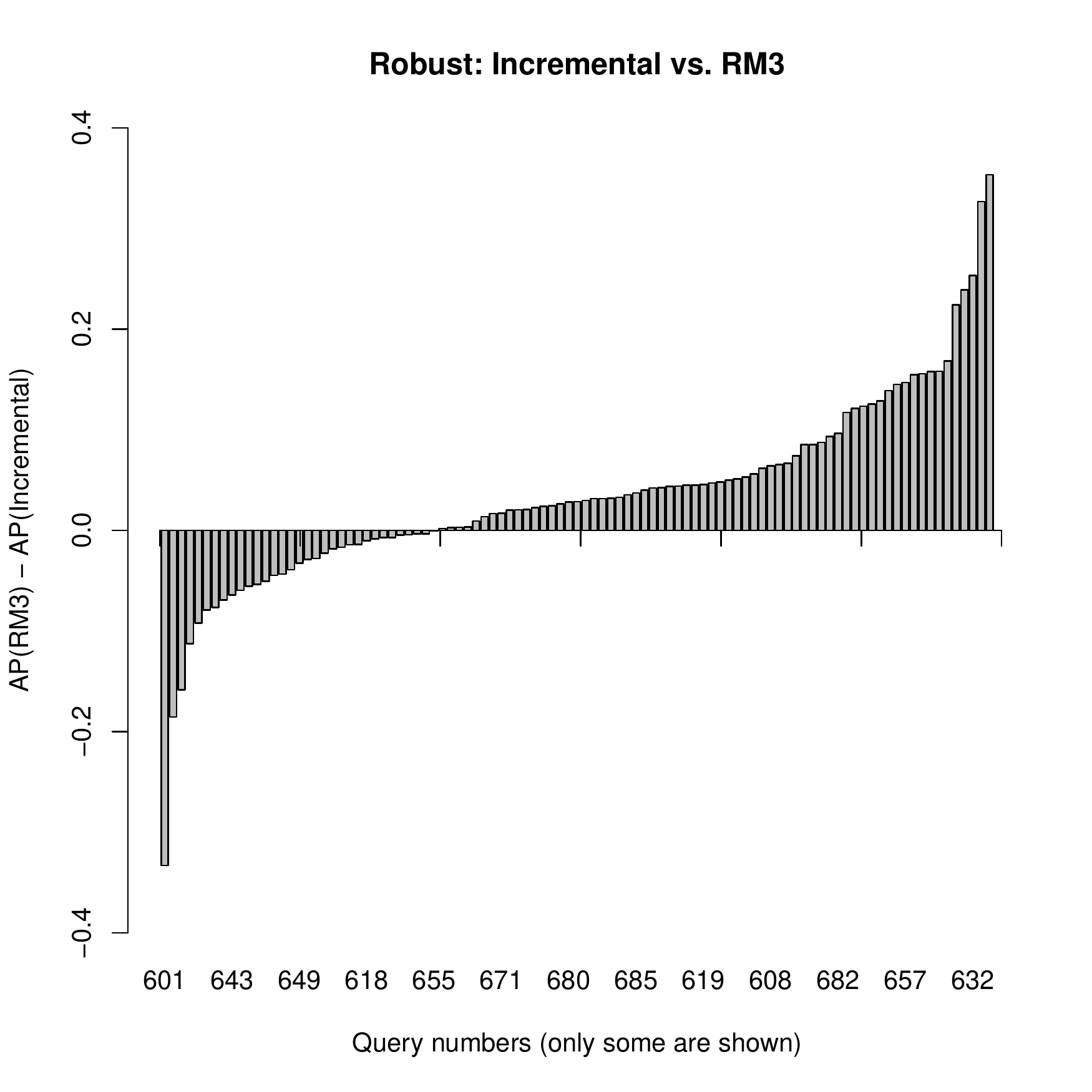}

  \includegraphics[scale=0.4]{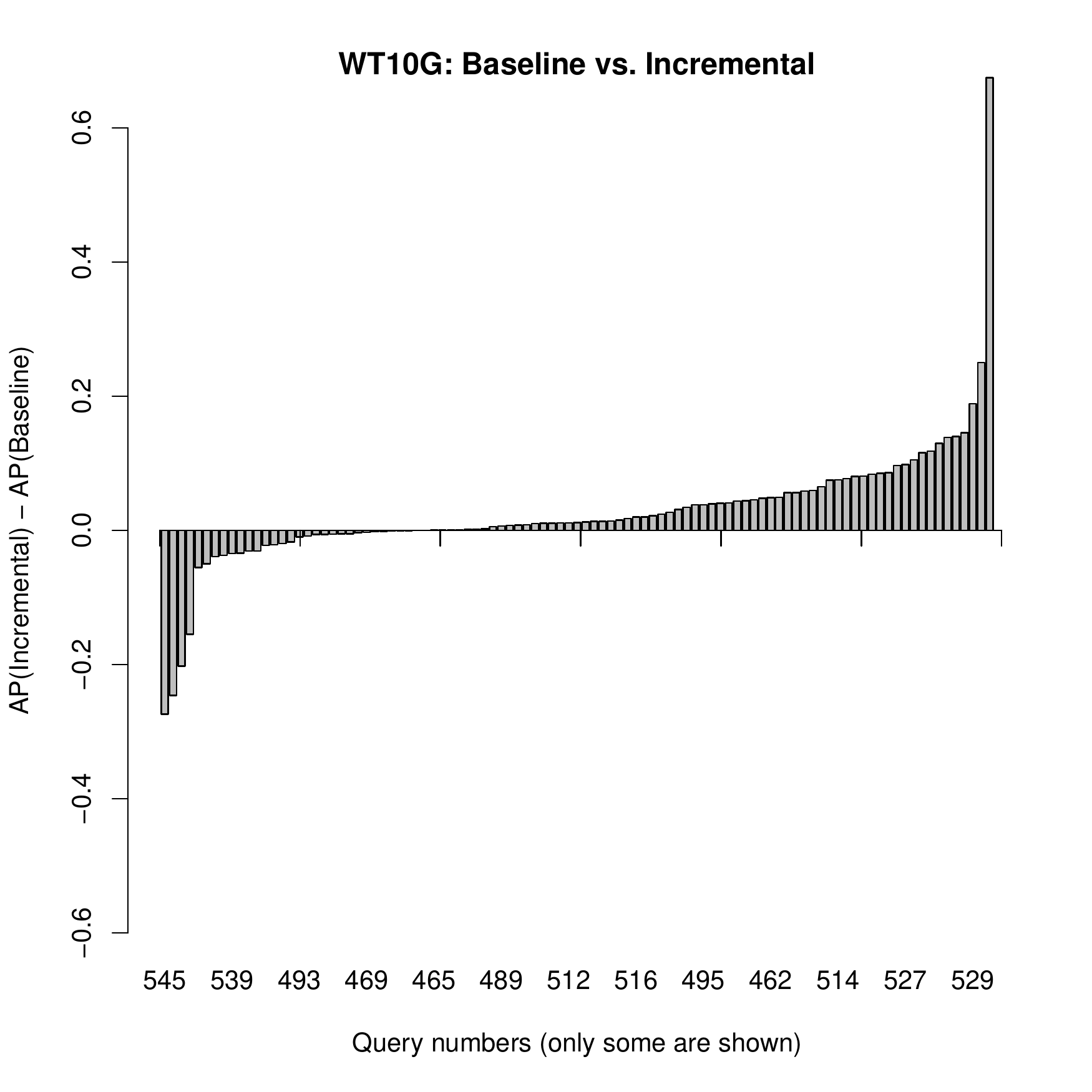}
  \hspace{1cm}
  \includegraphics[scale=0.4]{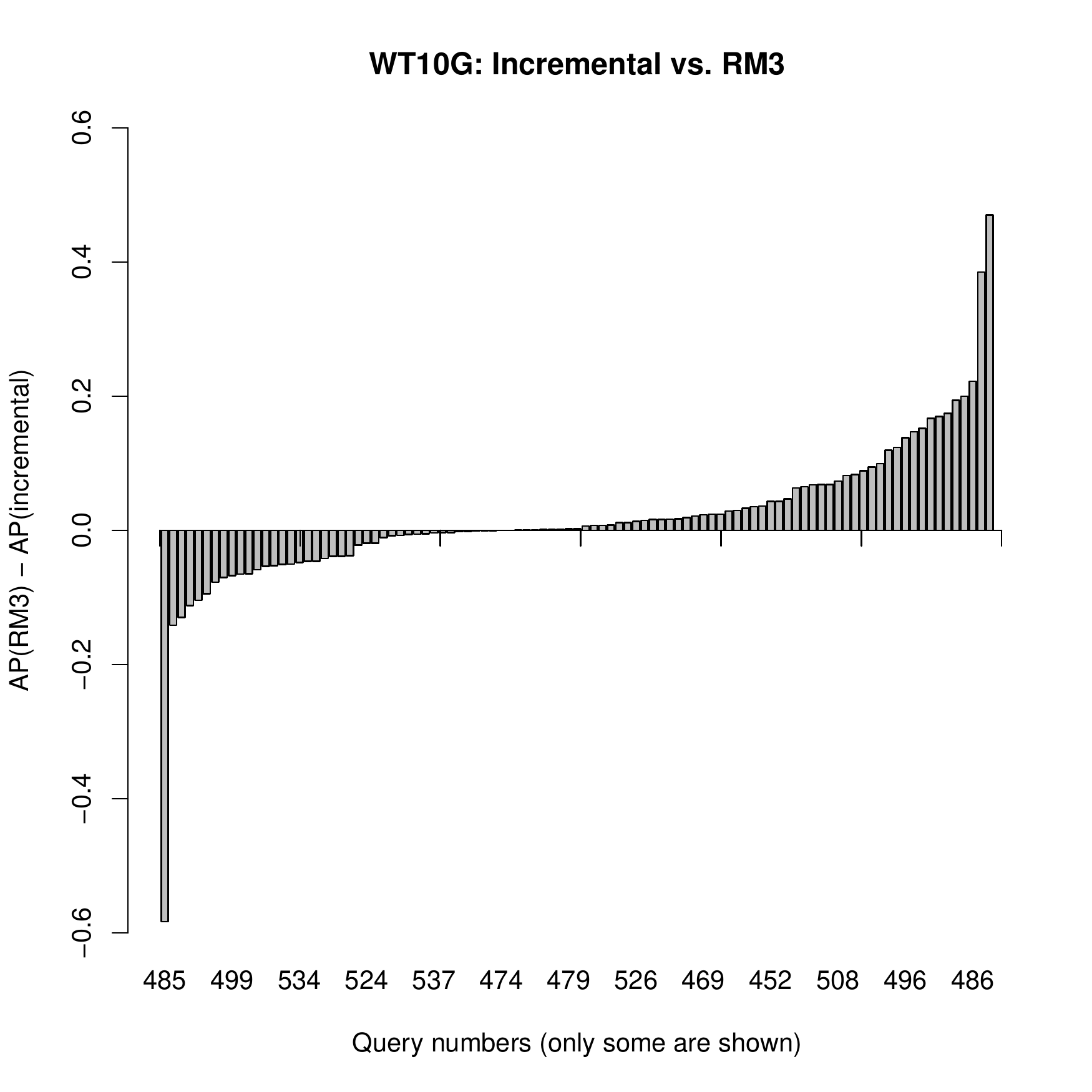}

  \caption{Difference in AP for individual queries.}
  \label{fig:qq}
\end{figure*}

A more detailed query-by-query comparison between the baseline, incremental
and RM3 methods is presented in Figure~\ref{fig:qq}. Each vertical bar in
the figure corresponds to a query, and the height of the bar is the
difference in AP for the two methods for that query. The figures show that,
as an expansion method, the incremental method is generally safe: it yields
improvements for most queries (bars above the X axis), and hurts
performance for only a few queries (bars below the X axis). However, RM3
``wins'' more often than it loses compared to the incremental method. While
these experiments provide some answers to questions 1 and 3 listed in the
Introduction, question 2 is harder to answer, and will require further
investigation.


\section{Discussion}
\label{sec:discuss}
Distributed neural language model word2vec, possesses the semantic and contextual information. 
This contributes to the performance improvement over text similarity based baseline for each of the three methods. 
Query expansion intuitively calls for finding terms which are similar to the query, and terms which occurs frequently in the relevant documents (captured from relevance feedback).
In the proposed embedding based QE techniques, the terms which are similar to the query terms in the collection-level abstract space are considered as the expansion terms.
Precisely, in the K-NN based QE method, expansion terms are chosen from the entire vocabulary, based on the similarity with query terms (or, composed query forms).
When the same K-NN based method is applied with feedback information, the search space is minimized, from the entire vocabulary, to the terms of top documents.
However the underlying similarity measure, that is the embedded vector similarity in the abstract space, remains the same.
This is the reason why K-NN and post-retrieval K-NN performs identically.
It is found that there is no significant difference between the performance between the two K-NN based QE methods\footnote{Using paired t-test with 95\% confidence measure.}.
However those techniques fails to capture the other features of potential expansion terms, such as terms, frequently co-occurring with query terms.
Experiments on the TREC ad-hoc and web datasets shows that the performance of RM3 is significantly better than the proposed methods which 
indicates that the co-occurrence statistics is more powerful than the similarity in the abstract space.

A drawback of the incremental KNN computation compared with post-retrieval KNN and pre-retrieval KNN QE is that the former takes more time, 
due to iterative pruning step involved.



\section{Conclusion and Future Work}
\label{sec:fw}
In this paper, we introduced some query expansion methods based on word embedding technique. 
Experiments on standard text collections show that the proposed methods are performing better than unexpanded baseline model.
However, they are significantly inferior than the feedback based expansion technique, such as RM3, which uses only co-occurrence based statistics to 
select terms and assign corresponding weights.
The obvious future work, in this direction, is to apply the embeddings in combination with co-occurrence based techniques (e.g. RM3).
In this work, we restrict the use of embeddings only to select similar words in the embedded space.
Thus a possible future scope is to use the embeddings exhaustively for utilizing other aspects of the embedded forms.
In our experiments, we trained the neural network over the entire vocabulary.
A possible future work is thus the investigation of local training of word2vec from pseudo-relevance documents which might get rid of the generalization effect when trained over the whole vocabulary.

\bibliographystyle{abbrv}
{
\bibliography{qe}

\begin{thebibliography}{10}

\bibitem{rm3}
N.~Abdul-Jaleel, J.~Allan, W.~B. Croft, O.~Diaz, L.~Larkey, X.~Li, M.~Smucker,
  and C.~Wade.
\newblock Umass at trec 2004: Novelty and hard.
\newblock In {\em Proc.\ TREC}, 2004.

\bibitem{deerwester1990indexing}
S.~Deerwester, S.~T. Dumais, G.~W. Furnas, T.~K. Landauer, and R.~Harshman.
\newblock Indexing by latent semantic analysis.
\newblock {\em JASIS}, 41(6):391, 1990.

\bibitem{GoodwinH14}
T.~Goodwin and S.~M. Harabagiu.
\newblock {UTD} at {TREC} 2014: Query expansion for clinical decision support.
\newblock In {\em Proc.\ {TREC} 2014}, 2014.

\bibitem{GrbovicDRSB15}
M.~Grbovic, N.~Djuric, V.~Radosavljevic, F.~Silvestri, and N.~Bhamidipati.
\newblock Context- and content-aware embeddings for query rewriting in
  sponsored search.
\newblock In {\em Proc.\ SIGIR 2015}, pages 383--392, 2015.

\bibitem{hofmann1999probabilistic}
T.~Hofmann.
\newblock Probabilistic latent semantic indexing.
\newblock In {\em Proc.\ SIGIR}, pages 50--57. ACM, 1999.

\bibitem{Mikolov13}
T.~Mikolov, I.~Sutskever, K.~Chen, G.~S. Corrado, and J.~Dean.
\newblock Distributed representations of words and phrases and their
  compositionality.
\newblock In {\em Proc.\ NIPS '13}, pages 3111--3119, 2013.

\bibitem{VulicM15}
I.~Vulic and M.~Moens.
\newblock Monolingual and cross-lingual information retrieval models based on
  (bilingual) word embeddings.
\newblock In {\em Proc.\ SIGIR '15}, pages 363--372, 2015.

\bibitem{weston2014memory}
J.~Weston, S.~Chopra, and A.~Bordes.
\newblock Memory networks.
\newblock {\em arXiv preprint arXiv:1410.3916}, 2014.

\bibitem{xu_query}
J.~Xu and W.~B. Croft.
\newblock Query expansion using local and global document analysis.
\newblock In {\em Proceedings of the 19th Annual International ACM SIGIR
  Conference on Research and Development in Information Retrieval}, SIGIR '96,
  pages 4--11, 1996.

\bibitem{zhai}
C.~Zhai and J.~Lafferty.
\newblock A study of smoothing methods for language models applied to
  information retrieval.
\newblock {\em ACM Trans. Inf. Syst.}, 2004.

\end{thebibliography}
}
\end{document}